# Simultaneous multi-patch-clamp and extracellular-array recordings: Single neuron reflects network activity

Roni Vardi[1,2,†], Amir Goldental[1,†], Shira Sardi[1], Anton Sheinin[3] and Ido Kanter[1,2,*]

[1]Department of Physics, Bar-Ilan University, Ramat-Gan 52900, Israel

[2]Gonda Interdisciplinary Brain Research Center and the Goodman Faculty of Life Sciences, Bar-Ilan University, Ramat-Gan 52900, Israel

[3]Department of Biochemistry and Molecular Biology, Tel Aviv University, Tel Aviv, Israel; Sagol School of Neuroscience, Tel Aviv University, Tel Aviv, Israel

†These authors contributed equally to this work.

*Correspondence: ido.kanter@biu.ac.il

**The increasing number of recording electrodes enhances the capability of capturing the network's cooperative activity, however, using too many monitors might alter the properties of the measured neural network and induce noise. Using a technique that merges simultaneous multi-patch-clamp and multi-electrode array recordings of neural networks in-vitro, we show that the membrane potential of a single neuron is a reliable and super-sensitive probe for monitoring such cooperative activities and their detailed rhythms. Specifically, the membrane potential and the spiking activity of a single neuron are either highly correlated or highly anti-correlated with the time-**

**dependent macroscopic activity of the entire network. This surprising observation also sheds light on the cooperative origin of neuronal burst in cultured networks. Our findings present an alternative flexible approach to the technique based on a massive tiling of networks by large-scale arrays of electrodes to monitor their activity.**

## Introduction

In a recent technology feature article "*Neurobiology: rethinking the electrode*"[1] it is stated that **"***No current technology can record from many thousands of individual neurons with single-cell resolution***"**. Fairly saying, nowadays, intracellular recording techniques are limited to several neurons only, where new types of extracellular in-vitro and in-vivo recordings are emerging with a moderate increase in the number of electrodes per array[2-5]. Nevertheless, the interplay between the spatial sampling resolution of the spiking activity of a neural network and its cooperative synchronized mode of activity is obscure[6]. Specifically, the *minimal number* and types of extracellular and intracellular electrodes required to capture accurately the momentary cooperative phenomena of large-scale neural networks, e.g. temporal synchronized modes or bursts, is unknown. On one hand, an increase in the number of recording electrodes increases the capability of capturing the network cooperative activity, however, at the moment we are technologically somehow limited[1]. On the other hand, the use of a large amount of monitors, i.e. electrodes, increase the sources of noise in the measured systems and might alter the properties of the measured neural network, e.g. neuronal plasticity[7,8]. Here we propose a different approach for studying dynamics of neural networks, where the macroscopic measure of the spiking activity and their rhythms are inferred from the membrane potential of a patched neuron. The effectiveness of this approach is surprising as it suggests that a single neuron, a single stochastic node[9] in a complex network with diverse properties, contains a reliable information on the dynamics of macroscopic properties of the entire network.

**Results**

**The Scheme of the Versatile Setup Combining Multi-Electrode-Array and Multi-Patch-Clamp Recordings.** Using a newly available setup, enabling extracellular recording from a multi-electrode array simultaneously with multi-patch-clamp[10,11] recordings (Figure 1 and Methods), we show that the membrane potential of a single neuron reflects extremely well the cooperative network spiking activities, including the timings of bursts and their instantaneous rhythms and patterns. This reflection is valid even in the case of a distant probe; where the location of the intracellularly recorded neuron is at a distance of several millimeters or even a centimeter away from the area sampled by the extracellular electrodes (Figure 2A).

Our in-vitro apparatus measurement (Figure 2A$_1$) consists of a 60-electrode array with a diameter of 30 micrometers each. The electrodes are separated by 0.5 mm from each other and cover an area of (2.5 mm) X (4.5 mm) (Figure 2A$_2$), consisting of around 2% of the entire ~5 cm$^2$ cortical tissue culture (gray circle in Figure 2A$_1$). The multi-electrode array samples the spontaneous firing activity of the neural network, consisting of around one million interconnected neurons[12] (see Methods). The unique capability of our system is to record from all the extracellular electrodes of the array simultaneously with intracellular recordings from patched neurons (demonstrated in Figure 2A). The recordings of the intra- and extra- cellular signals are done by two independent recording systems (Figure 1 and Methods) and require a careful synchronization of their clocks. A sustained 20 μs matching between the two clocks was achieved using careful analysis of the drift of the two clocks and by using leader-laggard triggers for synchronization (Supplementary Fig. S1).

**A Single Neuron Recording Reflects the Cooperative Network Burst Activities.** The raster plot of the activity recorded by the 60 extracellular electrodes over a period of ten minutes is exemplified in a snapshot of 100 seconds (Figure 2B, upper panel). The activity is governed by macroscopic cooperation among neurons comprising the network in the form of activity bursts[13-17], separated by periods of vanishing activity, and is quantitatively estimated by the temporal rate of the multi-electrode array (Figure 2B, middle panel, and Methods). The entire duration of a typical burst ranges between several dozens of milliseconds to several seconds and it is separated by long silent periods which can be extended to dozens of seconds (Figure 2B, upper panel)[18]. Each such burst consists of consecutive short bursts ranging between few dozens of milliseconds and hundreds of milliseconds that are separated by several dozens of milliseconds of vanishing activity (Figure 2C, upper panel). The large variability in the burst sizes and the durations of the inter burst intervals is related to the research of neuronal avalanches[19,20], which is beyond the scope of the current work.

The patch-clamp recording (Figure 2B, lower panel) of a neuron which is located close to one of the extracellular electrodes of the array (Figure 2A$_3$) indicates a complete correlation with the burst activity of the network. The patch-clamped neuron demonstrates a burst if and only if a spontaneous burst is recorded by the multi-electrode array. This similarity is extended even to the structure of the consecutive short bursts that constitute each "entire burst" of few seconds, on a timescale of dozens of milliseconds each (Figure 2C and Supplementary Fig. S3A). Hence, a single neuron reflects the cooperative spontaneous activity of the entire network on a timescale of seconds and even follows the fast erratic changes on a timescale of dozens of milliseconds. Note that during the entire network burst period the membrane potential of the patched neuron is evidently above the baseline membrane potential[21,22], ~-75 mV, (Figure 2C (lower panel), Figure 3 and Supplementary Fig. S2). This phenomenon originates from the massive EPSPs, which result from the cooperative activity of the network and serves as another evident marker for the network burst activities. Results are not limited to a specific location of the patched neuron, e.g. near the multi-electrode array, and were found to be robust to any other location as long as the neuron exhibits spontaneous activity. A similar correlation of activity between the extra- and intra-cellular recordings was found in dozens of spontaneously active cultures (Methods) as well as for patched neurons that were distant several millimeters out of the multi-electrode array.

The discussed correlation between the burst activity and the spiking activity and the membrane potential of the patched neuron is visually evident and its quantitative statistical description is given in Figure 3. The time differences between bursting activities and intracellularly recorded spikes (Figure 3A) indicate an average absolute delay of ~16 milliseconds between these two types of recording activities. This delay is typical but may vary among different cultures and as a function of the position of the patched neuron relative to the MEA. Another statistical measure is the conditional probability density function of the membrane potential with respect to the rate being above or below its average, as measured by the MEA (Figure 3B). Results clearly indicate that in between the bursts the membrane is mostly at its resting potential, where during bursts the membrane potential is significantly increased.

**Recordings of Two Distant Neurons Reflect the Cooperative Network Burst Activities.**

To examine whether the abovementioned correlation is extended to distant neurons, we measured the activity sampled by the multi-electrode array simultaneously with *two* patched neurons that were located several millimeters out of the multi-electrode array (Figure 4A). The correlation in the activities among two such patched neurons and a distant multi-electrode array strongly supports the generality of our main claim.

The position of the two patched neurons, relatively far from the multi-electrode array (Figure 4A), has to be compared to the previous case of one near patched neuron (Figure $2A_1$). These recordings were performed using a different neuronal culture, characterized by much longer silent periods between bursts with durations of several dozens of seconds (Figure 4B, upper panel). The two patched neurons follow this trend too; they demonstrate synchronized long silent periods between their burst activities (Figure 4B, middle and lower panels) which completely match the long silent periods measured by the multi-electrode array (Figure 4B, upper panel). Hence, the patched neurons reflect the network activity over dozens of seconds, where no burst was observed in their activity during the long silent periods of the network activity. A zoom-in into the first burst (Figure $4C_1$) and the last burst (Figure $4D_1$) indicates that there is a good correlation between the timings of the short bursts, composing the longer bursts, of the two patched neurons (Figures 4C$_1$ and 4D$_1$, middle and lower panels). In addition, there is a good fit between the timings of these short bursts and the short bursts recorded by the multi-electrode array (Figures 4C$_1$ and 4D$_1$). Note that during the entire network's burst period the membrane potentials of the patched neurons are evidently above the resting potential, ~-70 mV, (Figures 4C and 4D), which originates from the massive incoming synaptic currents to the neurons as a result of the cooperative activity of the network. Even in the case where the incoming currents generate several evoked spikes only (Figures 4C$_2$ and 4D$_2$, middle panels) the neuronal membrane potential functions as an evident marker for the background network burst activities. An additional zoom-in to the 400 milliseconds at the beginning of the first burst (Figure 4C$_2$) and the last burst (Figure 4D$_2$) indicates that there is a small time-shift of very few milliseconds between the first spike at the two patched neurons (Figure 4C$_2$ and 4D$_2$, middle and lower panels). Moreover, a comparison between the timings of the first spike of each of the patched neurons (or deviation from the baseline membrane potential, e.g. -70 mV) and the initiation of the network bursts measured by the multi-electrode array (e.g. the first moment where the temporal rate exceeds 5 Hz) indicates a time-shift of about 50 milliseconds. This time-shift might be attributed to the spatial distance of several millimeters between the center of the multi-electrode array and the location of the measured patched neurons. Since the spatial location of the beginning of a spontaneous burst is not fixed, the activity of the network can either first come across the patched neurons (Figure 4C) or the multi-electrode array (Figure 4D). Nevertheless, these time-shifts (up to dozens of milliseconds) are short compared with the timescales of the entire bursts (seconds) and the time-lags in between them (up to dozens of seconds) (Figure 4B), hence the synchronous activities can be easily detected.

**An anti-correlated neuronal temporal state.** In a limited number of patched neurons, a different feature of activity was observed, characterized by an anti-correlation between the firing of the patched neuron and the network activity during each "entire burst" (Figure 5). Specifically, an "entire burst" lasts for a few seconds and is composed of several shorter bursts (see also Figure 2C). The patched neuron fires only in between the shorter bursts, thus demonstrating an anti-correlation. During the longer silent periods between "entire bursts", the activity of the patched neuron can continue or decay. Note that the temporal firing pattern of the patched neuron is similar to other patched neurons and would be interpreted, without the multi-electrode array recording, as aligned with the network bursts (Figure 2). This rare phenomenon was revealed only when comparing simultaneously the temporal recordings of the multi-electrode array and the patch-clamp, demonstrating the significance of these combined techniques.

## Discussion

The observation that the dynamics of the membrane potential of a patched neuron reflects the activity of the entire macroscopic neural network, presents an alternative flexible technique to the technique based on a massive tiling of the network by fixed positions and structures of a large-scale array of extracellular electrodes. Our findings, obtained from cortical tissue cultures, indicate that bursts of a single patched neuron and its membrane potential reflect the activity of the entire macroscopic network, together with its internal details and with a very precise time resolution. Although it is possible to find an isolated evoked spike of the patched neuron with the lack of similar activity in the network (Supplementary Fig. S3B), it is very rare to find such a counterexample of a burst in a network that does not translate into the neuronal membrane potential. In addition, our preliminary results indicate that adding synaptic blockers to a spontaneous active culture[23] (Methods) prevents both bursts measured by the multi-electrode array and by the patched neuron, while the rinse of the blockers revealed bursts simultaneously both in the patched neuron and in the multi-electrode array (Supplementary Fig. S4).

This technique also shed light on the controversial issue of the origin of neuronal bursts; whether it is a standalone internal neuronal feature, or it is a consequence of a cooperative synchronized activity of a large interconnected network, resulting in a strong long-lasting incoming current to a neuron, reflected by its membrane potential. Our results strongly support the latter scenario where it is very rare to observe a neuronal burst that is not accompanied by a nearby cooperative burst.

Another interesting phenomenon observed in some rare neurons using the presented method is an anti-correlation between the multi-electrode and the patch recorded spikes (Figure 5). The origination of this anti-correlated firing, observed independently of the distance of the patched neurons from the center of the MEA, might be a result of a strong inhibition of the presynaptic neurons of the patched neuron, as the network activity is correlated with the membrane potential but with the lack of firing. Nevertheless, the understanding of this phenomenon requires further research. Observing and studying this kind of phenomena requires a technique merging simultaneously multi-patch-clamp and multi-electrode array recordings of a large network, as presented in this study.

It is possible that other types of correlative structures exist between the activities of the network and the patched neuron, depending on the detailed structure of the network, e.g. connectivity, synaptic strengths and the detailed spatial arrangements of the inhibitory synapses. The fact that we only observed two types of correlative structures, correlations and anti-correlations, might be an outcome of the type of measured cultures or an outcome of the conditions, including nutrient supply and age of cultures and we cannot exclude the existence of a small fraction of isolated neurons that are not participating in the collective activity. It is possible that in other types of cultures or in-vivo networks, additional complex collective phenomena will be discovered, and their possible variety certainly deserves a further research. In the case of in-vivo systems, it was found that an up-state of a neuron is correlated with some macroscopic measurements[24], which is in agreement with the reported results. The current findings have been obtained in certain neural networks and under specific conditions only (dense cultures composed of cortical neurons) and our work calls to examine their validity on other types of neural networks. Exploring the relation between the reported phenomena and in-vivo observations may shed light on the origin of macroscopic and microscopic dynamics of neural activity within the human brain.

The disappearance and reappearance of synchronized bursts on the levels of a single neuron and the entire macroscopic network also strongly support the hypothesis that neuronal bursts are driven mainly by the cooperative network activity, as opposed to internal neuronal mechanisms. The power of the presented technique was exemplified in the case of simultaneous recordings from multi-patched neurons and multi-electrode extracellular array, however combining intra- and extra- cellular stimulations with such measurements is expected to further enhance the effectiveness and usefulness of the technique.

# Methods

**Animals**. All procedures were in accordance with the National Institutes of Health Guide for the Care and Use of Laboratory Animals and Bar-Ilan University Guidelines for the Use and Care of Laboratory Animals in Research and were approved and supervised by the Bar-Ilan University Animal Care and Use Committee.

**Culture preparation**. Cortical neurons were obtained from newborn rats (Sprague-Dawley) within 48 h after birth using mechanical and enzymatic procedures. The cortical tissue was digested enzymatically with 0.05% trypsin solution in phosphate-buffered saline (Dulbecco's PBS) free of calcium and magnesium, and supplemented with 20 mM glucose, at 37°C. Enzyme treatment was terminated using heat-inactivated horse serum, and cells were then mechanically dissociated. The neurons were plated directly onto substrate-integrated multi-electrode arrays (MEAs) and allowed to develop functionally and structurally mature networks over a time period of 2-4 weeks in vitro, prior to the experiments. The number of plated neurons in a typical network was in the order of 1,300,000, covering an area of about ~5 cm$^2$. The preparations were bathed in minimal essential medium (MEM-Earle, Earle's Salt Base without L-Glutamine) supplemented with heat-inactivated horse serum (5%), B27 supplement (2%), glutamine (0.5 mM), glucose (20 mM), and gentamicin (10 g/ml), and maintained in an atmosphere of 37°C, 5% $CO_2$ and 95% air in an incubator.

**Synaptic blockers.** Additional experiments were conducted on cultured cortical neurons that were functionally isolated from their network by a pharmacological block of glutamatergic and GABAergic synapses. For each culture 20 µl of a cocktail of synaptic blockers were used, consisting of 10 µM CNQX (6-cyano-7-nitroquinoxaline-2,3-dione), 80 µM APV (DL-2-amino-5-phosphonovaleric acid) and 5 µM Bicuculline methiodide. This cocktail did not necessarily block completely the spontaneous network activity, but rather made it sparse (Supplementary Fig. S1). At least one hour was allowed for stabilization of the effect.

**Stimulation and recording – MEA.** An array of 60 Ti/Au/TiN extracellular electrodes, 30 µm in diameter, and spaced 500 µm from each other (Multi-Channel Systems, Reutlingen, Germany) was used. The insulation layer (silicon nitride) was pre-treated with polyethyleneimine (0.01% in 0.1 M Borate buffer solution). A commercial setup (MEA2100-60-headstage, MEA2100-interface board, MCS, Reutlingen, Germany) for recording and analyzing data from 60-electrode MEAs was used, with integrated data acquisition from 60 MEA electrodes and 4 additional analog channels, integrated filter amplifier and 3-channel current or voltage stimulus generator. Each channel was sampled at a frequency of 50k samples/s.

**Stimulation and recording – Patch Clamp.** The Electrophysiological recordings were performed in whole cell configuration utilizing a Multiclamp 700B patch clamp amplifier (Molecular Devices, Foster City, CA). The cells were constantly perfused with the slow flow of extracellular solution consisting of (mM): NaCl 140, KCl 3, CaCl2 2, MgCl2 1, HEPES 10 (Sigma-Aldrich Corp. Rehovot, Israel), supplemented with 2 mg/ml glucose (Sigma-Aldrich Corp. Rehovot, Israel), pH 7.3, osmolarity adjusted to 300-305 mOsm. The patch pipettes had resistances of 3–5 MOhm after filling with a solution containing (in mM): KCl 135, HEPES 10, glucose 5, MgATP 2, GTP 0.5 (Sigma-Aldrich Corp. Rehovot, Israel), pH 7.3, osmolarity adjusted to 285-290 mOsm. After obtaining the giga-ohm seal, the membrane was ruptured and the cells were subjected to fast current clamp by injecting an appropriate amount of current in order to adjust the membrane potential to about -70 mV. The changes in neuronal membrane potential were acquired through a Dgidata 1550 analog/digital converter using pClamp 10 electrophysiological software (Molecular Devices, Foster City, CA). The acquisition started upon receiving the TTL trigger from MEA setup. The signals were filtered at 10 kHz and digitized at 50 kHz.

**Data analysis**. Analyses were performed in a Matlab environment (MathWorks, Natwick, MA, USA). The reported results were confirmed based on dozens of spontaneously active neural cultures where in each culture around eight patched neurons were examined. The recorded data **from the MEA** (voltage) was filtered by convolution with a Gaussian that has a standard deviation (STD) of 0.1 ms, and the threshold for action potential detection**, for the raster plot,** was defined to be 6 times the STD of this convolution.

The temporal rate is calculated as following:

$$r(t) = \left(\frac{1}{60}\sum_{t^*} \delta(t - t^*)\right) * f(t)$$

where *t* is the relevant time, the sum is over all spike times (*t*\*) recorded by the MEA, the asterisk symbols convolution and *f(t)* is a normalized Gaussian around 0 with a STD of 2 ms. The 1/60 is for averaging the rate per electrode such that the average value of *r(t)* is the average number of recorded spikes per second per electrode.

The distribution of time differences in Figure 3 was produced by finding the shortest time difference between each threshold crossing of the rate to above 20 Hz to a threshold crossing of the membrane potential to above -25 mV.

**Figures**

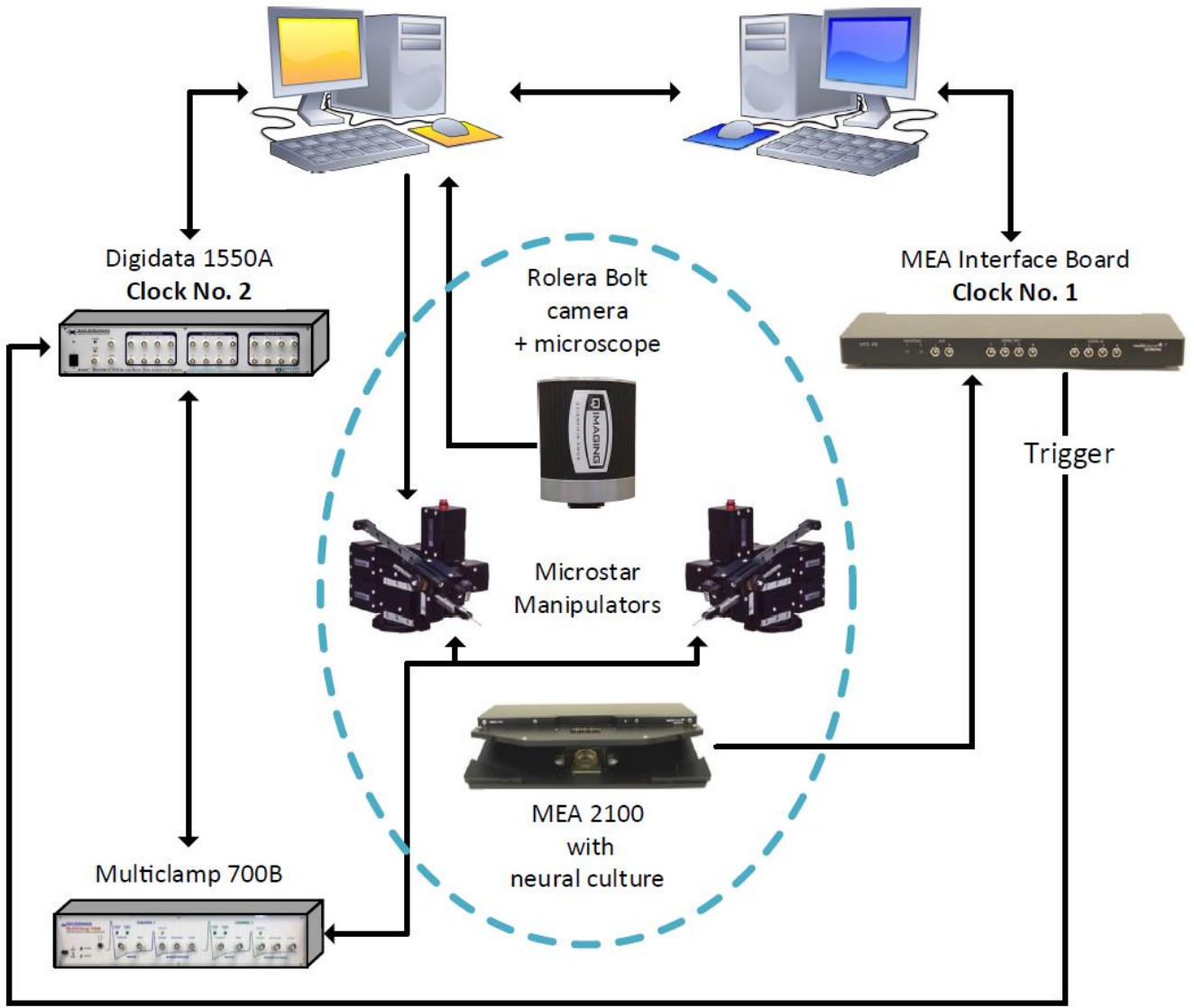

**Figure 1.** The scheme of the experimental setup combining multi-electrode array and patch recordings. The multi-electrode array, MEA 2100 (Methods), is controlled by the MEA interface boarded and a computer (blue, right computer). The Patch clamp sub-system consists of several microstar manipulators, an upright microscope (Slicescope-pro-6000, Sceintifica), and a camera. Stimulations and recordings are implemented using

multiclamp 700B and Digidata 1550A and are controlled by a computer (yellow, left computer). The recorded MEA/patch data is saved on the blue/yellow computer, respectively. The time of the MEA system is controlled by a clock placed in the MEA interface board (clock No. 1) and the time of the patch subsystem is controlled by a clock placed in the Digidata 1550A (clock No. 2). The relative timings are controlled by triggers sent from the MEA interface board to the Digidata using leader-laggard configuration.

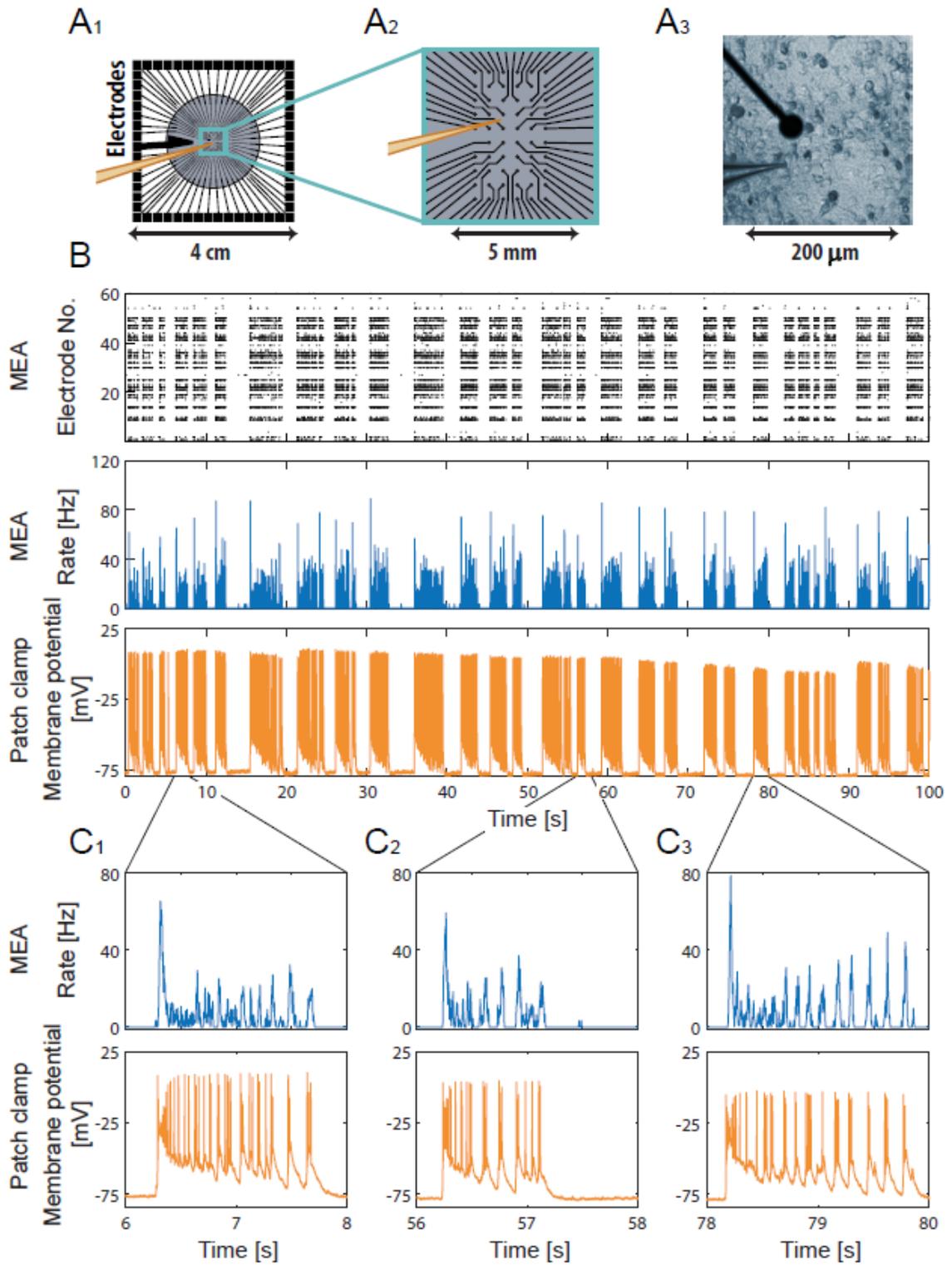

**Figure 2.** A single neuron current-clamp recording reflects the cooperative network burst activities recorded by an extracellular multi-electrode array (MEA). (**A$_1$**) A schema of the in-vitro apparatus measurement consisting of a 60 multi-electrode array. The cortical tissue culture (~1.3 million neurons) is plated in the gray circle area, ~5 cm$^2$. The light-blue square represents the area covered by the extracellular electrodes. The orange electrode represents an intracellular patch electrode. (**A$_2$**) A zoom-in of the area covered by the extracellular electrodes. Extracellular electrodes are separated by 0.5 mm with a diameter of 30 μm each, covering an area of (2.5 mm) X (4.5 mm). (**A$_3$**) A snapshot of a part of the neuronal culture, demonstrating simultaneous recordings with an extracellular electrode (top) and an intracellular patch electrode (bottom). (**B**) Upper panel: A raster plot of the activity recorded by the 60 extracellular electrodes over a period of 100 seconds. Each row represents the activity recorded by an extracellular electrode and each dot represents a detected spike. Middle panel: The temporal rate activity of the multi-electrode array presented in the upper panel was calculated using a convolution (see Methods). Lower panel: The membrane potential of the current-clamped neuron placed near an extracellular electrode, (A$_3$), recorded simultaneously with the extracellular recordings shown in the upper panels. (**C**) A zoom-in of three of the bursts shown in (B) (middle and bottom panels), for the network firing rate, recorded by the extracellular electrodes (top), and the voltage of the single neuron recorded by an intracellular electrode (bottom). Results indicate a high degree of correlation between the burst activities of the network and the single neuron.

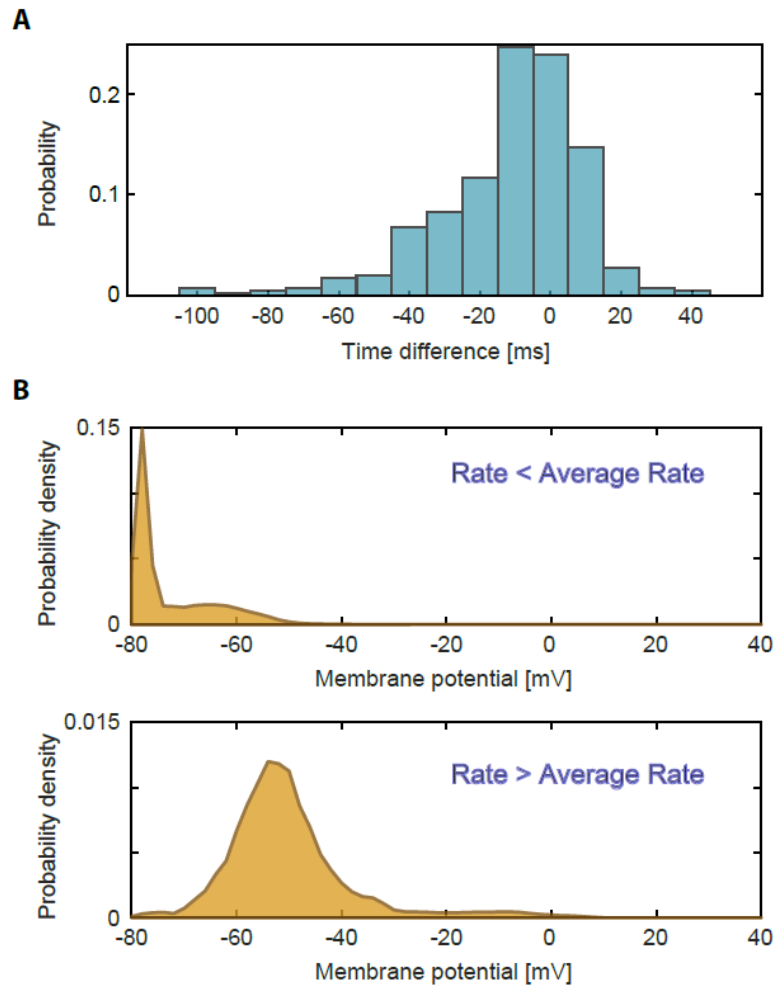

**Figure 3.** Quantitative statistical description of the correlations between bursting activities and intracellular spiking and membrane potential recordings. (**A**) The probabilities of the time difference between the intracellularly recorded spikes and MEA bursts of Figure 2. Specifically, for every threshold crossing of the rate to above 20 Hz, the shortest time difference to a threshold crossing of the neuronal membrane potential to above -25 mV was calculated. (**B**) The probability density function of the membrane

potential of the neuron of Figure 2, separated for sampling points where the temporal rate of the MEA recorded spikes is above its average, ~2.8 Hz, (upper panel) and below its average (lower panel). The sum up of these two probability densities together is unity.

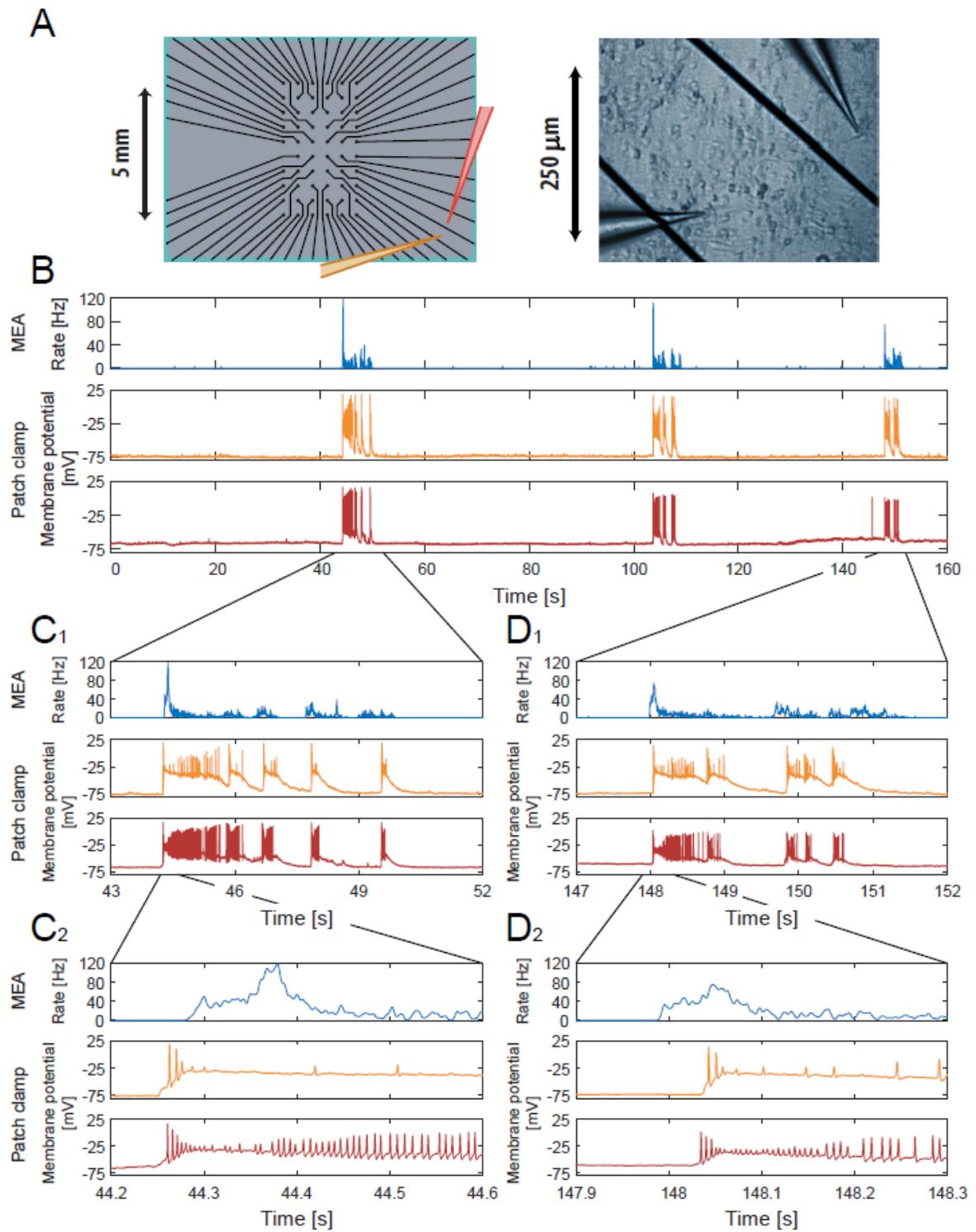

**Figure 4.** Extracellular multi-electrode array (MEA) recordings simultaneous with two distant neurons intracellular current-clamp recordings, indicating high correlation

between cooperative burst of the network and single neuron membrane potential. (**A**) Left: A zoom-in of the area covered by the extracellular electrodes, similar to Figure 1A$_2$. The orange and red electrodes represent two intracellular patch electrodes placed several millimeters out of the multi-electrode array. Right: A snapshot of part of the neuronal culture together with two neurons with two patch electrodes. (**B**) Similar to Figure 1B. Upper panel: The rate activity of the multi-electrode array, over a period of 160 seconds, calculated using a convolution (see Methods). Middle and lower panels: The membrane potential of two current-clamped neurons placed several millimeters out of the multi-electrode array, (A), recorded simultaneously with the extracellular recordings shown in the upper panel. (**C$_1$**) A zoom-in of the first burst shown in (B), for the network firing rate, recorded by the extracellular electrodes (top), and the voltage of the two patched neurons recorded by the intracellular electrodes (middle and bottom). (**C$_2$**) An additional zoom-in on the 400 milliseconds at the beginning of the burst shown in (C$_1$). (**D$_{1-2}$**) Similar to (C$_{1-2}$), for the last burst shown in (B). Results indicate a high degree of correlation between the extracellular recorded cooperative network activity and the two recorded membrane potentials.

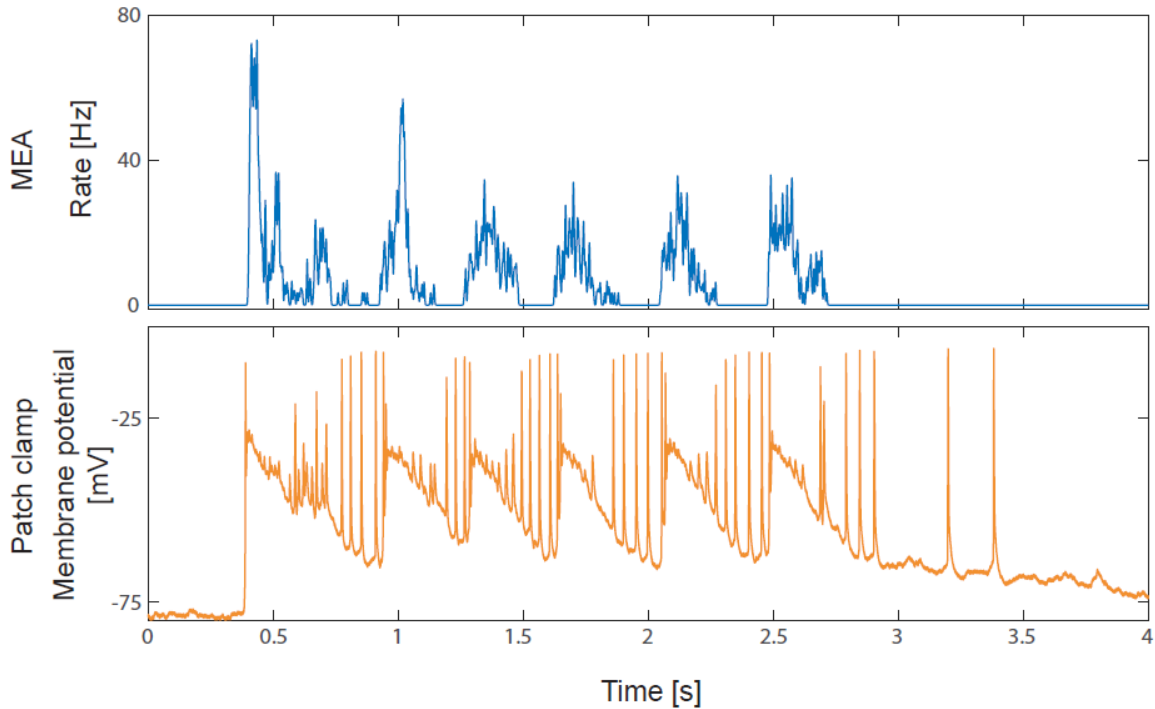

**Figure 5.** Anti-correlation between extra- and intra- cellular recordings within the burst. An example of an entire burst of the network (upper panel) recorded by the multi-electrode array and the patch-clamp membrane potential (lower panel). The network burst lasts for about 2 seconds and is composed of several shorter bursts separated by silent periods of no activity of the network. The membrane potential of the patched neuron is above its baseline membrane potential (~-75 mV) during the entire burst of the network, however, the evoked spikes occur mainly during the silence periods of the network. We hypothesize this type of anti-correlation indicates another feature where a single neuron reflects the network cooperative activity, originates from inhibition, but its understanding requires further research.

## Acknowledgements

We wish to thank Moshe Abeles for helpful discussions and Pinhas Sabo for technical assistance. This work was supported by the TELEM grant of the planning and budgeting Committee of the Council of Higher Education, 2015, Israel.

## Author contributions

R.V. and S.S. prepared the tissue cultures. R.V. performed the extra- and intra- cellular experiments with the help of A.S.. A.S. and A.G. helped in the operation and the installation of the new combined MEA and patch system. R.V. and A.G. analyzed the data. A.G. developed the numerical package to analyze the data. I.K. initiated the study and supervised all aspects of the work. All authors discussed the results and commented on the manuscript.

**Supplementary Information**

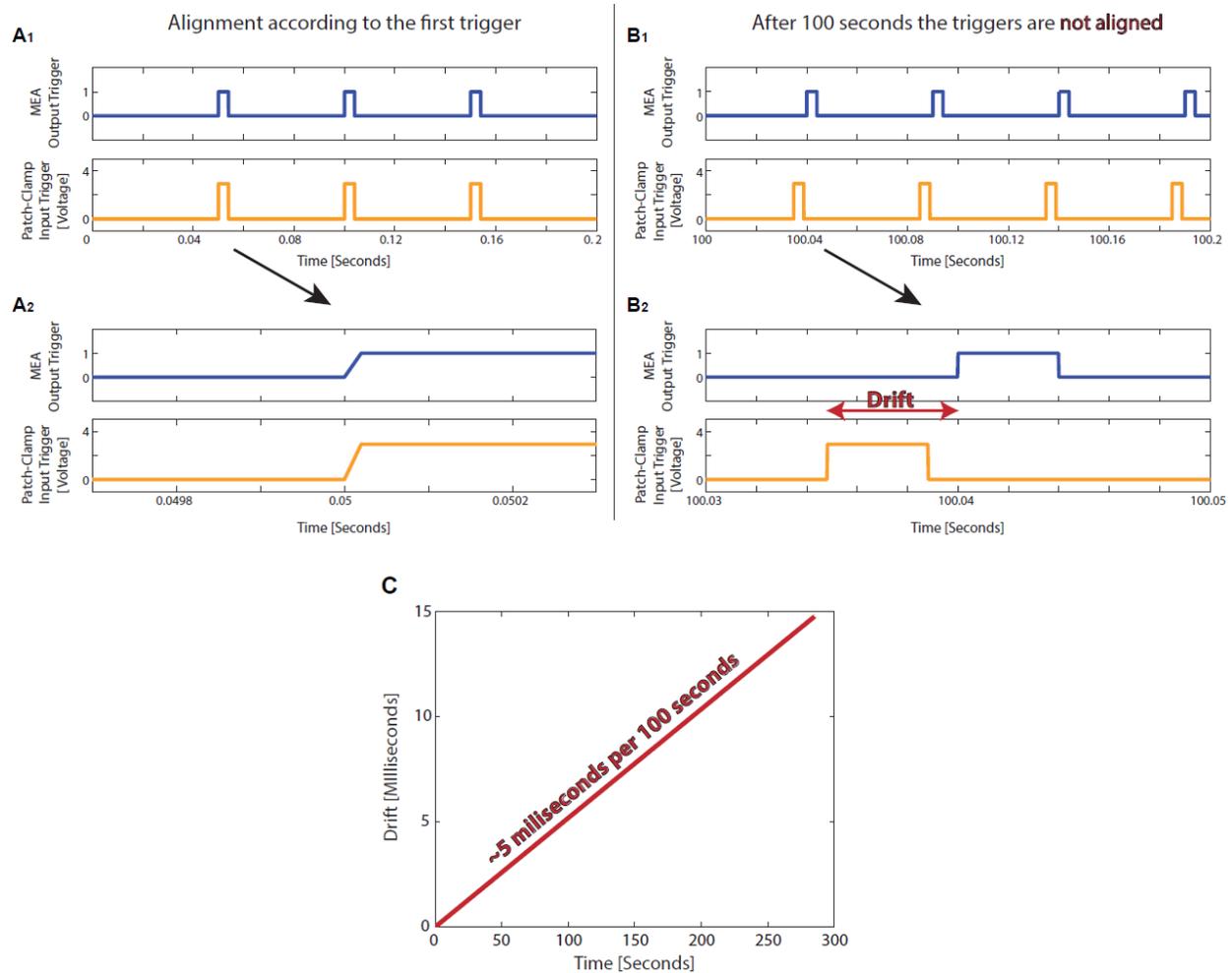

**Figure S1. The drift between the two clocks.** The synchronization between the two clocks is implemented using triggers, sent in a leader-laggard configuration (Figure 1 in the manuscript) from the MEA to the patch subsystem (blue curves in panels A and B). The output trigger is sent every 50 milliseconds from the MEA and its duration is 4 ms (however only the rise timing of the pulse is important and has a duration of 20 microseconds). The recorded input trigger to the patch subsystem is 3.8 V and its duration is 4 ms too (orange curves in panels A and B). Initially, the recoded data by the MEA and patch clamp systems are aligned through the first trigger (panels $A_1$ and $A_2$). The drift between the clocks after

~100 seconds is visible (panels $B_1$ and $B_2$) and is estimated quantitatively (panel C). This drift was carefully balanced and canceled out in the presented data.

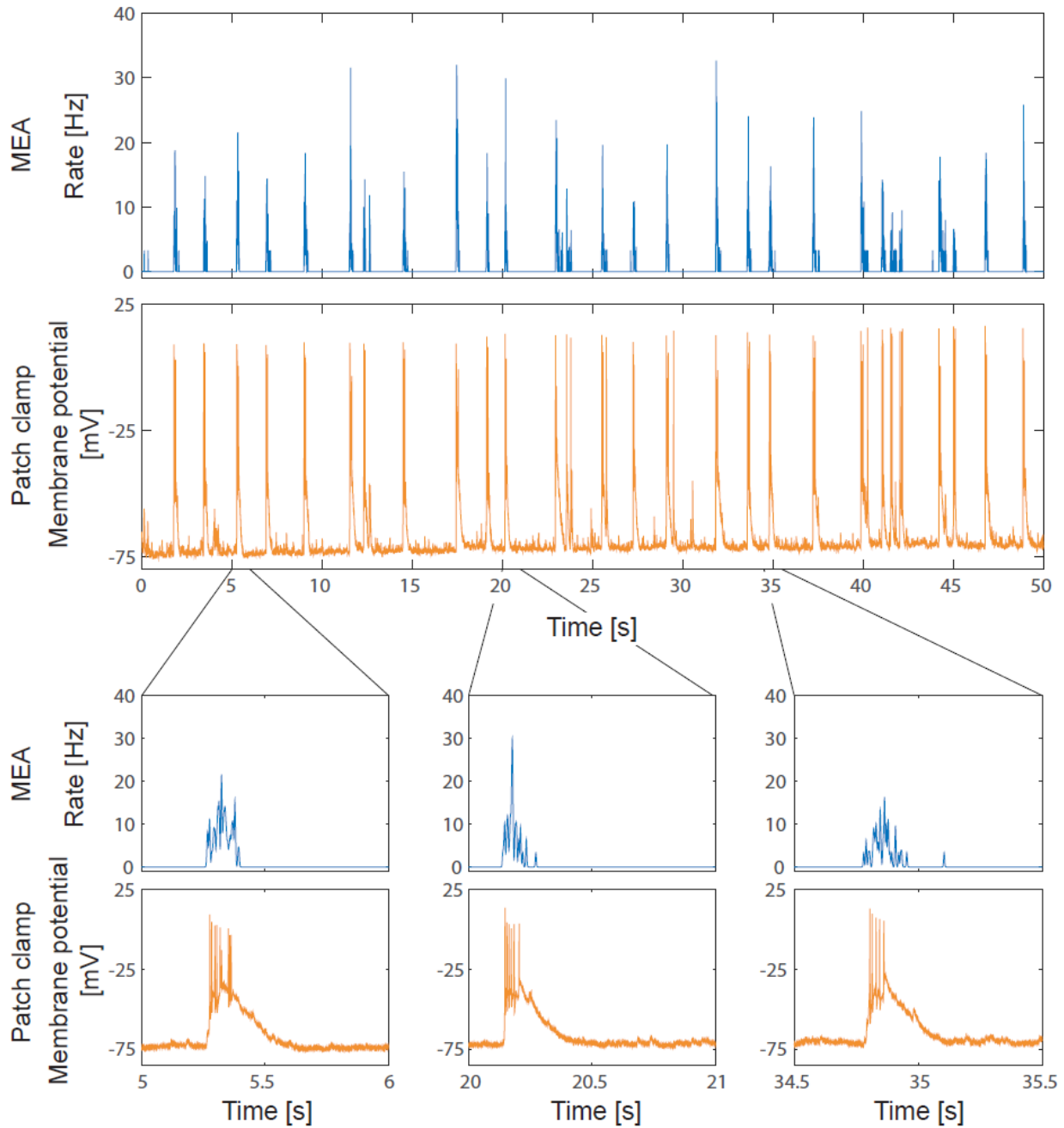

**Figure S2. Another example of 50 seconds recording.** Similar to Figure 2 in the manuscript, indicating that a single neuron current-clamp recording reflects the cooperative network burst activities recorded by an extracellular multi-electrode array (MEA). Upper panel: The temporal rate activity of the multi-electrode array. Middle panel: The membrane potential of the current-clamped neuron indicating high correlation with the MEA recordings. Lower

panels: Zoom-in of three of the bursts shown in the upper panels. Results indicate high correlation between the bursts of the network and the single neuron internal dynamics, with a short time shift of less than 20 milliseconds.

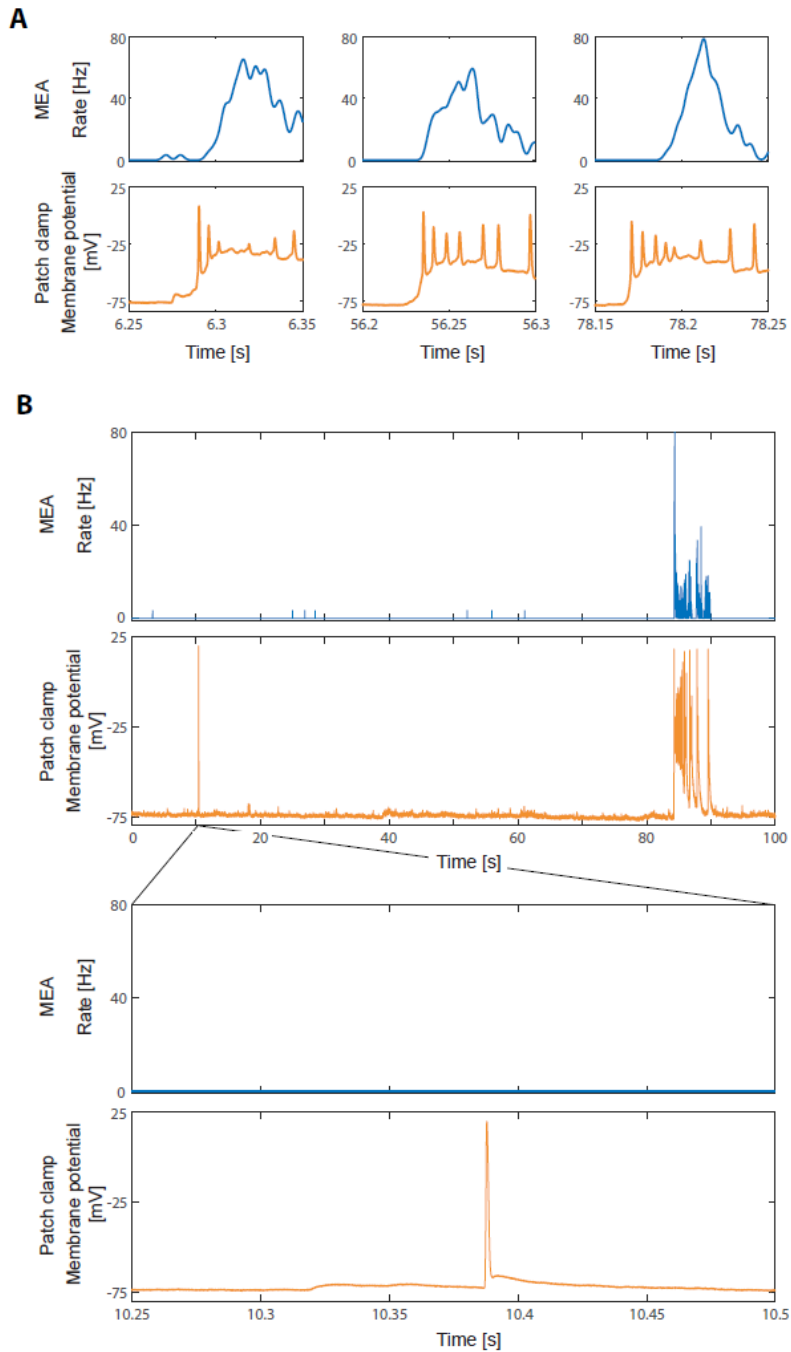

**Figure S3. The correlation between the network's activity and the neuronal membrane potential.**
A: A zoom-in of Figure 2C in the manuscript. A zoom-in of three of the bursts shown in Figure 2C in the manuscript for the network firing rate, recorded by the extracellular electrodes (top), and the membrane potential of the single neuron recorded by an intracellular electrode (bottom). Results indicate a high correlation between the bursts of the network and the single neuron. A comparison between the timing of the first spike of the patched neuron (or deviation from the baseline membrane potential voltage, e.g. -75 mV) and the initiation of the network bursts measured by the multi-electrode array (e.g. the first moment where the temporal rate exceeds 5 Hz) indicates a time-shift of very few milliseconds in the left and the middle panels, and about 50 milliseconds in the right panel. B: A single spike recoded by the patch clamp system that does not necessarily reflect a burst of the network. An example where a network burst is accompanied by a burst of the patched neuron (upper two panels). Nevertheless, it is possible to find an isolated evoked spike of the patched neuron with the lack of similar activity in the network (zoom-in of the relevant time-slot is shown in the two lower panels).

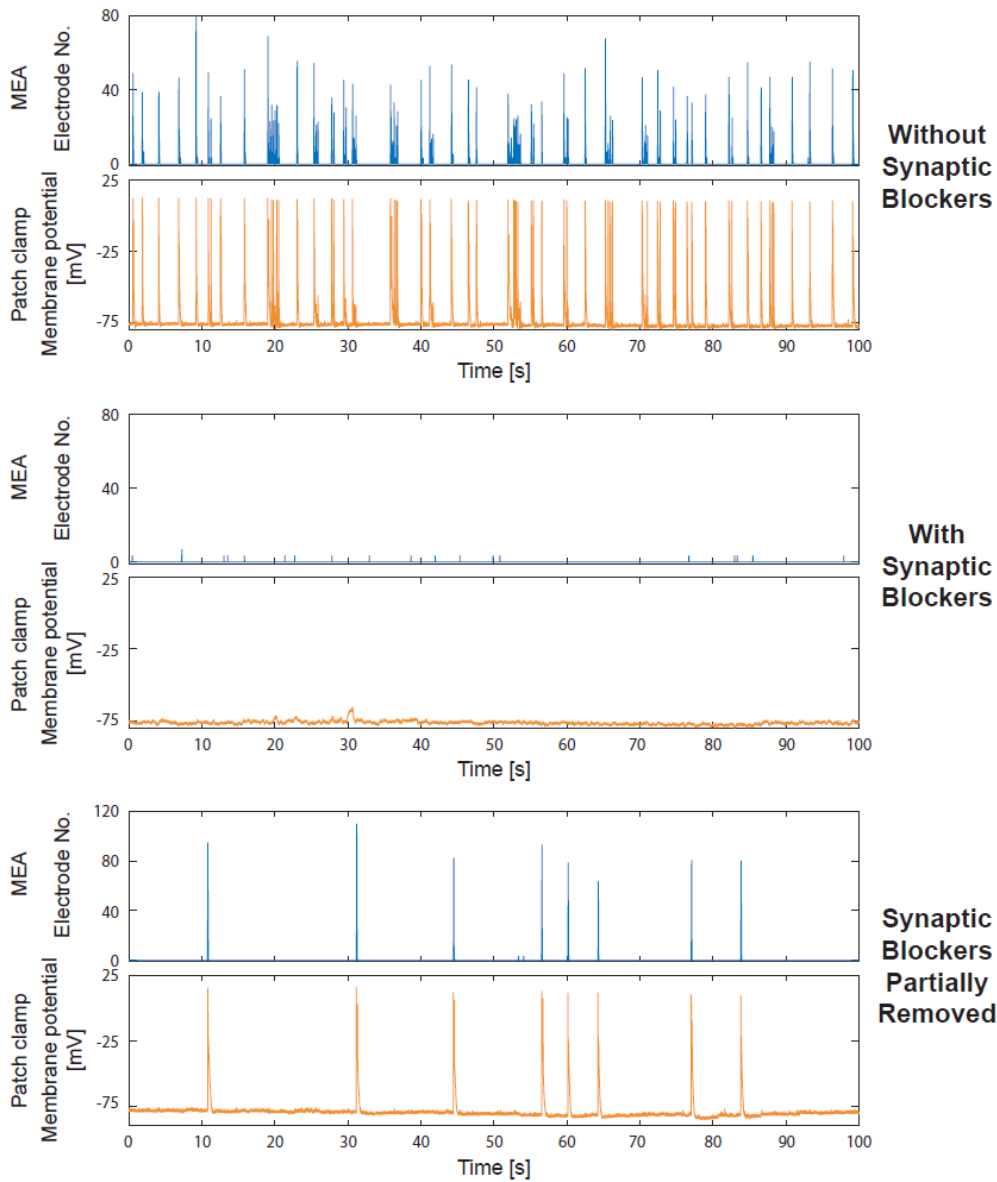

**Figure S4. The robustness of the result that a single neuron reflects network activity even under pharmacological manipulations.** Upper two panels: The extra-cellular rate activity, recorded simultaneously with the membrane potential of the current-clamped neuron, similar to middle and lower panels of Figure 1B in the manuscript. A high correlation between the timings of the network and the neuronal bursts is evident. Middle panels: When adding synaptic blockers (Online Methods), the cooperative bursts of the network almost vanish as well as the neuronal bursts and postsynaptic potentials. Lower panels: The

partial rinse of the blockers revealed short bursts simultaneously both in the patched neuron and in the multi-electrode array.